\documentclass{aa}  

\usepackage{graphicx}
\usepackage{txfonts}
\begin{document} 

   \title{Orbital and sub-orbital period determination of the candidate high-mass X-ray binary HD 3191}

%   \subtitle{I. Overviewing the $\kappa$-mechanism}

\author{Josep Mart\'{\i}\inst{1}
\and
Pedro L. Luque-Escamilla\inst{2}
\and
Estrella S\'anchez-Ayaso\inst{3}
\and
Josep M. Paredes\inst{4}
}

\institute{
Departamento de F\'isica (EPSJ), Universidad de Ja\'en, Campus Las Lagunillas s/n, A3, E-23071  Ja\'en, Spain\\
\email{jmarti@ujaen.es}
\and 
Departamento de Ingenier\'{\i}a Mec\'anica y Minera (EPSJ), Universidad de Ja\'en, Campus Las Lagunillas s/n, A3, E-23071 Ja\'en, Spain\\
\email{peter@ujaen.es}
\and
Departamento de Ciencias Integradas, Centro de Estudios Avanzados en F\'{\i}sica, Matem\'atica y Computaci\'on, Universidad de Huelva, E-21071, Huelva, Spain\\
%Departamento de Ciencias Integradas, Facultad de Ciencias Experimentales, Campus "El Carmen", Universidad de Huelva,
 %Avda. de las Fuerzas Armadas s/n, E-21007  Huelva, Spain\\
 \email{estrella.sanchez@dci.uhu.es}
\and
Departament de F\'{\i}sica Qu\`antica i Astrof\'{\i}sica,  Institut de Ci\`encies del Cosmos, Universitat de Barcelona, IEEC-UB, Mart\'{\i} i Franqu\`es 1, E-08028 Barcelona, Spain\\
%\email{jmparedes@ub.edu}
 }

   \date{Received XXXXXX  XX, XXXX; accepted XXXXXX  XX, XXXX}

% \abstract{}{}{}{}{} 
% 5 {} token are mandatory
 
  \abstract
   {}
   {The final aim of this paper is to expand the sparse group of X-ray binaries with gamma-ray counterparts as laboratories to study high-energy processes
   under physical conditions that periodically repeat.}
   {A  follow-up of a candidate system has been carried out. We applied both photometric and spectroscopic techniques in the optical domain together with a period analysis 
   %mainly
   using  the phase dispersion minimisation and CLEAN methods. A tentative period search was also conducted in the gamma-ray domain.}
   {Our main result is having established the binary nature of the optical star and X-ray emitter HD 3191 towards the {\it Fermi} gamma-ray source
   4FGL J0035.8+6131, the last one proposed to be associated with a blazar of an unknown type. An orbital period close to 16 days is reported for HD 3191 together with a 
   likely rotation, or pulsation, period of about 0.6 d.
   Although no convincing
   evidence for the orbital cycle has been found in the {\it Fermi} light curve up to now,
   the confirmed presence of a high-mass X-ray binary towards 4FGL J0035.8+6131 now strengthens the need for caution about its true nature.}
   {}


   \maketitle
%
%-------------------------------------------------------------------

\section{Introduction}

Attention was drawn to the bright and luminous star HD 3191 (spectral type B1 IV:nn)  a few years ago after the detection
of the bright gamma-ray transient {\it Fermi} J0035+6131 \citep{2016ATel.8554....1P}. XMM-{\it Newton} observations of the {\it Fermi} flaring source
location (0.08$^{\circ}$ radius at 95\% confidence) soon identified two main X-ray candidates consistent with it: 
the radio source 87GB 003232.7+611352 and
the luminous star HD 3191 \citep{2016ATel.8783....1P}. Follow-up studies point to the 87GB source,
also identified as LQAC 008+061 \citep{2015A&A...583A..75S},
 as the most likely counterpart
candidate, which is believed to be an active galaxy or quasar  serendipi\-tously located behind the Galactic disk \citep{2018ApJ...862...83P}. However, these authors could not completely rule out
the possibility of HD 3191 being associated with the {\it Fermi} emitter. 
The parallax measurement recently provided by the Gaia Early Data Release 3 (EDR3) places this star at a distance of $1.21 \pm 0.02$ kpc \citep{2020arXiv201201533G}.
In any case, the star's  unabsorbed hard X-ray emission ($L_X \simeq 2.6 \times 10^{31}$ erg s$^{-1}$, based on the \citet{2018ApJ...862...83P} flux and Gaia distance)
 strongly points towards a new, but almost quiescent, high-mass X-ray binary system. Optical spectroscopy of HD 3191 by \cite{2016ATel.8789....1M},
 with resolving power of $\lambda/\Delta \lambda \sim 11000$, detected no remarkable peculiarity in its spectrum.
The only exception was a radial velocity of $-46.0 \pm 0.5$ km s$^{-1}$ that, intriguingly, did not match an old previous value  of $-22 \pm 3$ km s$^{-1}$  in the literature
\citep{1961PDAO...12....1P}. On the other hand, the Gaia radial velocity %in the Gaia Data Release 2  
is quoted as $+72 \pm 12$ km s$^{-1}$.
These three different velocities are suggestive of a binary system. However,  the highly discrepant Gaia value needs to be considered with caution as it was derived
using a template spectrum too cold (6500 K) for an early-type star.
Nevertheless, other clues about the possible binary nature of HD 3191 come from its proposed runaway behaviour according to
\citet{2011MNRAS.410..190T}, who analysed its {\it Hipparcos} astrometry. A slightly higher proper motion is reported in the recent Gaia EDR3, thus rendering
it more conceivable that HD 3191 experienced  a kick due to a supernova explosion in binary system in the past. Solving this issue is still in need of a better radial velocity determination.

The Large Area Telescope (LAT) on board the orbiting {\it Fermi} gamma-ray observatory recently provided new insights on the transient source J0035+6131.
 The LAT 10 year source catalogue\footnote{https://fermi.gsfc.nasa.gov/ssc/data/access/lat/10yr\_catalog/} in its second data release (4FGL-DR2)  
 includes it as the source 4FGL J0035.8+6131.
The accompanying light curve shows two additional epochs of faint detection in addition to the two flaring intervals
 analysed by \cite{2018ApJ...862...83P}. This suggests that this GeV source could remain active at faint levels instead of a strict transient behaviour. 
 Nevertheless, the nature of the source still remains open as the 4FGL-DR2 catalogue still classifies it as a blazar of an unknown type. 
 Gamma-ray and soft gamma-ray flaring activity has also been suspected from different Galactic high-mass X-ray binaries, 
 such as the black hole candidate MWC 656 \citep{2014Natur.505..378C}  or the well-known gamma-ray binary LS I +61 303 \citep{2009A&A...497..457M}.

In this context, we have devoted some effort to investigating the nature of the secondary counterpart candidate to {\it Fermi} J0035+6131/4FGL J0035.8+6131,
namely the star HD 3191, whose possible association with the gamma-ray source is still conceivable.  
Even if unrelated to the {\it Fermi} emission, this early-type star is an interesting object by itself in
any case.
The following sections contain an account of our observational work, in the optical domain, mostly based on our own ground-based observations and the
analysis of archival data from space. As a main result, the HD 3191 binary nature is confirmed with an accurate estimate of the orbital  
%and rotational
period.
With this knowledge in hand, the possibility of using the matching period approach is now open to help test the HD 3191 candidacy.
Our preliminary search for the signature of the system period in the {\it Fermi} data  yields, however, non-conclusive results.

%Var index 26.31155

\section{Spectroscopic and photometric observations from the ground}

\begin{figure}[t]
\includegraphics[width=7.5cm, angle=-90]{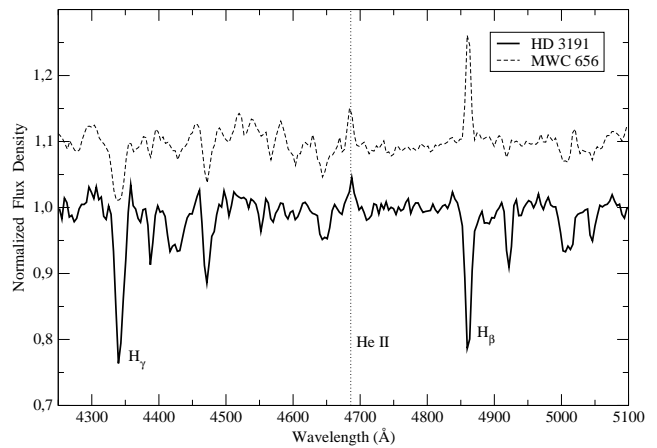}
%\imageii
\caption{   \label{LISA} Spectrum of HD 3191 obtained from OUJA on 2019 December 16 and clearly showing the detection of the He II emission line, whose
4686 $\AA$ wavelength is marked by the pointed vertical line. The dashed spectrum is shown for comparison purposes and corresponds to the Be black hole system MWC 656.
It was  observed with the same instrumental setup on 2019 November 18. Balmer lines in the spectral region are also indicated.
}%% no full stop at the end
\end{figure}

We started our observational study of HD 3191 at the Observatory of the University of Ja\'en \citep[OUJA,][]{2017BlgAJ..26...91M},
 whose 41 cm telescope was equipped in late 2019 with a low-resolution LISA spectrograph from {\it Shelyak Instruments} ($\lambda/\Delta \lambda \sim1100$),
 combined with a Peltier cooled ATIK 460EX camera.
 To our surprise, we noticed the clear presence of an emission line of ionised helium (He II) at the 4686 $\AA$ wavelength as illustrated in Fig. \ref{LISA}.
 This feature in emission is more common in O-type stars than in B-type stars, thus its presence was intriguing. Moreover,
 this finding had not been previously reported by the spectroscopic observers mentioned above. All of this stimulated our interest to continue our
 HD 3191 observations. The emission feature was repeatedly detected during the following weeks with different degrees of intensity, which are to be reported elsewhere.
 Its equivalent width, when most prominently detected in Fig. \ref{LISA}, amounted to about $-0.54$ $\AA$.

 Given that our spectroscopic data lack enough resolution for appropriate radial velocity work, we soon concentrated our efforts on acquiring frequent photometric data.
 The LISA instrumental setup also allowed us to perform unfiltered photometry by using its secondary guiding camera (Atik 314L+) whose CCD sensor (ICX285AL)
 has a spectral sensitivity that peaks in approximate coincidence with the  Johnson $V$-band. A simple zero-point was applied to the resulting photometry
 to approximate the $V$-band magnitudes. We determined it on a photometric night
 from reference stars in the open cluster NGC 7790, just a few degrees away from the target. The OUJA photometric data points over 40 nights are reported in Table 
 \ref{phot} as a function of heliocentric Julian date (HJD). 
 The check stars within the HD 3191 field of view are not included here, but they remained stable typically within $\pm0.01$ magnitudes.

\begin{figure}
\includegraphics[width=7.0cm, angle=-90]{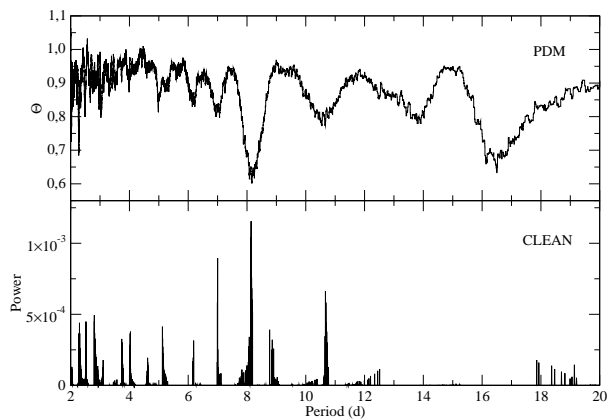}
%\imageii
\caption{   \label{ouja}  PDM (top) and CLEAN (bottom) periodograms of OUJA photometry in Table \ref{phot}.
}%% no full stop at the end
\end{figure}

\begin{figure}
\includegraphics[width=7.0cm, angle=-90]{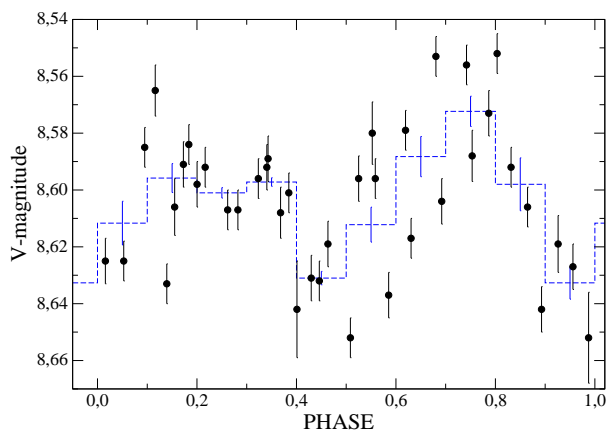}
%\imageii
\caption{   \label{folded_ouja}  OUJA photometric observations folded on an orbital period of 16.28 days as estimated using CLEAN. The blue dashed line
indicates the average with bins of one-tenth the cycle.
}%% no full stop at the end
\end{figure}

Periodograms of Table \ref{phot} photometry were computed using the
 phase dispersion minimisation  \citep[PDM,][]{1978ApJ...224..953S}
and CLEAN \citep{1987AJ.....93..968R} methods. The result is displayed in Fig. \ref{ouja} where both methods point to the existence of a significant period
at $8.19 \pm 0.06$ days and $8.14\pm 0.05$ days, respectively. 
We soon suspected this to be the signature of an orbital modulation of an ellipsoidal type, as expected
from a candidate X-ray binary. If this was the case, then the double-wave symmetry of the ellipsoidal modulation would imply that the true orbital period has to be doubled.
The OUJA phase folded light curve is shown in Fig. \ref{folded_ouja} using twice the best CLEAN period estimate (16.28 days), where hints of this modulation are present. 
At this stage, phase zero was preliminarily set at HJD 2458749.5 for plotting purposes. 
All  spectroscopic and photometric OUJA frames were processed using IRAF tasks and scripts \citep{1993ASPC...52..173T}.

 % 
%-------------------------------------------------------------
%                                             Simple A&A Table
%-------------------------------------------------------------
%
\begin{table}
\caption{Results of OUJA differential photometry$^*$ of HD 3191.}             % title of Table
\label{phot}      % is used to refer this table in the text
\centering                          % used for centering table
\begin{tabular}{c c c c}        % centered columns (4 columns)
\hline\hline                 % inserts double horizontal lines
HJD & $V$-mag & HJD  & $V$-mag \\    % table heading 
$-2450000.0$  &  & $-2450000.0$ &  \\
\hline                        % inserts single horizontal line
     8856.272085 &  8.596  $\pm$ 0.007  &  8888.286860   & 8.596   $\pm$ 0.008 \\
    8857.261180 &  8.579  $\pm$ 0.007  &  8889.269872   & 8.637   $\pm$ 0.008 \\
    8858.256316 &  8.553  $\pm$ 0.007  &  8893.276992   & 8.592   $\pm$ 0.007 \\
    8859.257958 &  8.556  $\pm$ 0.007  &  8894.274682   & 8.642   $\pm$ 0.008 \\
    8860.260147 &  8.552  $\pm$ 0.007  &  8895.302680   & 8.627   $\pm$ 0.008 \\
    8861.256622 &  8.606  $\pm$ 0.007  &  8896.276793   & 8.625   $\pm$ 0.008 \\
    8862.255585 &  8.619  $\pm$ 0.010  &  8898.288181   & 8.633   $\pm$ 0.007 \\
    8863.253625 &  8.652  $\pm$ 0.016  &  8899.278413   & 8.598   $\pm$ 0.008 \\
    8864.317974 &  8.625  $\pm$ 0.007  &  8900.280749   & 8.607   $\pm$ 0.007 \\
    8865.349406 &  8.565  $\pm$ 0.009  &  8901.282955   & 8.596   $\pm$ 0.007 \\
    8866.271078 &  8.591  $\pm$ 0.008  &  8902.281687   & 8.601   $\pm$ 0.007 \\
    8876.267014 &  8.573  $\pm$ 0.008  &  8903.280076   & 8.632   $\pm$ 0.007 \\
    8881.284466 &  8.585  $\pm$ 0.007  &  8904.297556   & 8.652   $\pm$ 0.007 \\
    8882.264924 &  8.606  $\pm$ 0.010  &  8906.282994   & 8.617   $\pm$ 0.007 \\
    8883.263655 &  8.592  $\pm$ 0.007  &  8907.286113   & 8.604   $\pm$ 0.008 \\
    8884.335476 &  8.607  $\pm$ 0.007  &  8908.282320   & 8.588   $\pm$ 0.009 \\
    8885.281738 &  8.592  $\pm$ 0.008  &  8915.288567   & 8.584   $\pm$ 0.007 \\
    8885.322513 &  8.589  $\pm$ 0.008  &  8918.291054   & 8.608   $\pm$ 0.009 \\
    8886.267990 &  8.642  $\pm$ 0.017  &  8919.295372   & 8.631   $\pm$ 0.008 \\
    8887.282586 &  8.619  $\pm$ 0.008  &  8921.292557   & 8.580   $\pm$ 0.011 \\
\hline                                   %inserts single line
\end{tabular}
\tablefoottext{*}{Errors quoted do not include the possible systematic uncertainty ($\sim 0.1$ mag) from the adopted zero point value.}
\end{table}

\section{Photometric observations from  space}
 
 The light curve in Fig. \ref{folded_ouja} shows a possible double-wave ellipsoidal effect, but with a significant dispersion that could cast doubt on the
 reality of the orbital period estimated above.
 This fact prompted us to try to improve the quality of our analysis. At this point,  we noticed the availability of HD 3191 in the data releases
 of the {\it Transiting Exoplanet Survey Satellite} ({\it TESS}\footnote{{\tt https://tess.mit.edu}}). The quality of {\it TESS} photometry is unprecedented compared to ground based
 telescopes such as ours. Moreover, it provides a continuous time coverage of several weeks in different sections of the sky and with a time cadence of 2 minutes.
 Its wide-field cameras have a 600-1000 nm bandpass, centred on the Cousins $I_C$-band at a wavelength of 7865 $\AA$. 
 
\begin{figure}
\includegraphics[width=9.5cm, angle=0]{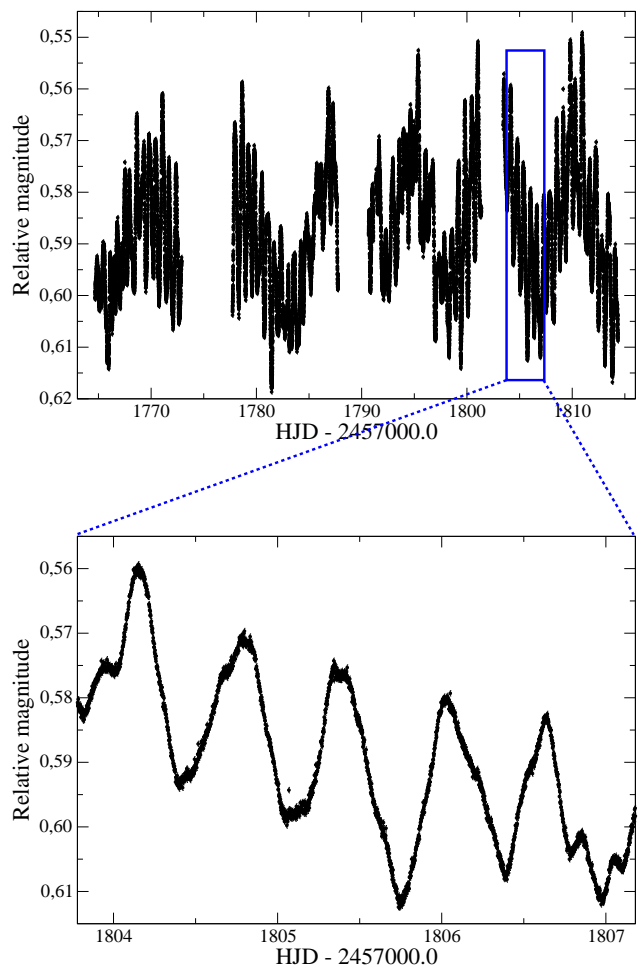}
%\imageii
\caption{   \label{tess_lc}  {\bf Top.} Light curve of HD 3191 based on PDC data from the 
TESS space observatory (Sectors 17 and 18). {\bf Bottom}. Zoomed view of a representative sample of TESS data to illustrate 
how its time cadence of two minutes enables us to clearly see intra-day variability.
}%% no full stop at the end
\end{figure}

 We retrieved  all the presearch data conditioning (PDC) data points  of HD 3191 from the {\it TESS} archive. 
 This includes data from Sectors 17 and 18 obtained during 2019. PDC data take different instrumental corrections into account, such as the focus, pointing,
 discontinuities due to occasional radiation events, as well as the identification and removal
 of outlier points. Light-loss and star crowding problems inside the photometric aperture
 are also dealt with \citep{2016SPIE.9913E..3EJ}. Nevertheless, this last issue is not a severe problem for HD 3191 since optical stars in its arc-minute vicinity are 
 several magnitudes fainter and none are reported as a variable in the SIMBAD database.
 Finally, no long-term trend was removed from the PDC files.
 The corresponding light curve is displayed in the top panel of Fig. \ref{tess_lc}, where both long-term (days) and short-term (hours) variability is evident.
 Brightness changes on intra-day timescales are better noticed in the representative zoomed section in the bottom panel of the figure.

\begin{figure}
\includegraphics[width=7.5cm, angle=-90]{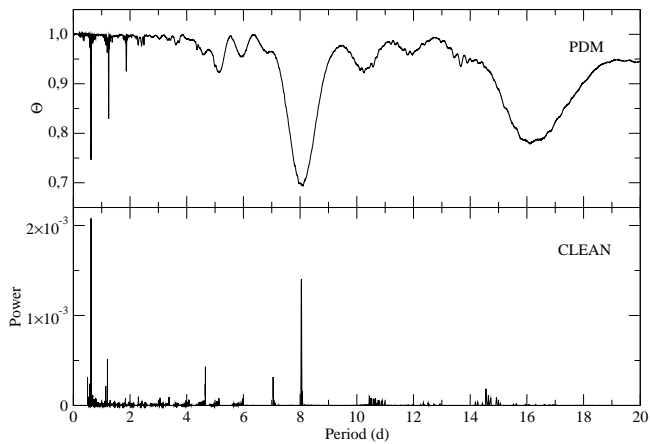}
%\imageii
\caption{   \label{tess_per}  PDM (top) and CLEAN (bottom) periodograms of TESS photometry (Sectors 17 and 18).
}%% no full stop at the end
\end{figure}

\subsection{Period searches}

A periodicity analysis using again PDM and CLEAN is presented in Fig. \ref{tess_per}. Here, a (doubled) period very close to sixteen days could be immediately 
 confirmed and, at the same time, improved with respect to OUJA data. The (doubled) period value provided by the two algorithms 
  turned out to be very consistent as they yielded $16.11 \pm 0.01$ d and $16.09 \pm 0.01$ d for PDM and CLEAN, respectively.
  Moreover, both periodograms also unveiled a second significant period related to the short-term variations. These also turned out to repeat with a period
  estimated as  $0.62296 \pm 0.00002$ d (PDM) and $0.62295 \pm 0.00002$ d (CLEAN), respectively.

Hereafter, we adopt the TESS CLEAN values of the two periods found  for discussion purposes. 
PDM works fine with non-sinusoidal variations but, as shown below, this is not the case here. 
The phase origin is the same for both periods, but it has been  set to HJD 2458750.14, which is justified later.

\subsection{Period stability}

\begin{figure}
\includegraphics[width=8.0cm, angle=0]{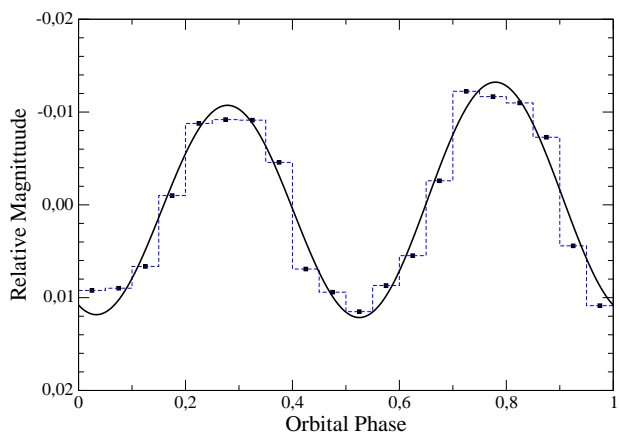}
%\imageii
\caption{   \label{qlporb}  Comparison of the modulation due to 0.06215 and 0.12430 day$^{-1}$ sine waves with the FFI photometry of HD 3191 from {\it TESS}
Sectors 17, 18, and 24 folded on the 16.09 day period and averaged using 0.05 phase bins.
The phase origin adopted is HJD 2458750.14.
}%% no full stop at the end
\end{figure}

\begin{figure}
\includegraphics[width=8.0cm, angle=0]{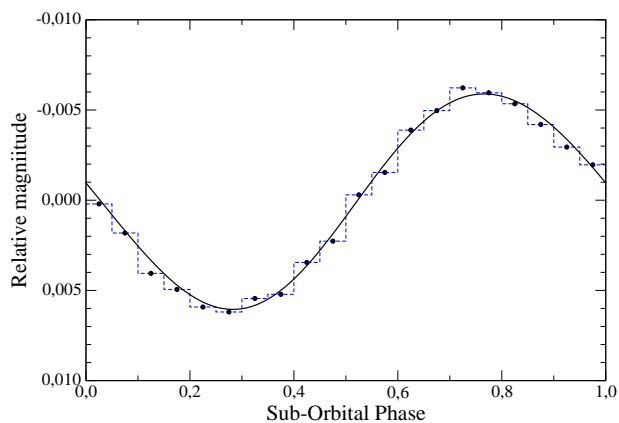}
%\imageii
\caption{   \label{qlprot} 
Comparison of the modulation due to 1.60527 and 3.21053  day$^{-1}$ sine waves with the FFI photometry of HD 3191 from {\it TESS}
Sectors 17, 18, and 24 folded on the 0.62295 day period and averaged using 0.05 phase bins.
The same zero phase as in the previous figure has been arbitrarily adopted.}
%}%% no full stop at the end
\end{figure}

In an attempt to assess period stabilities over timescales as long as possible, we explored additional {\it TESS} data  acquired during 2020 from Sector 24.
HD 3191 was also covered then, although with a time sampling of 0.5 h. 
To assemble a homogeneous dataset, we used the {\it lightkurve} software \citep{2018ascl.soft12013L} to process the TESS full frame image (FFI) files
for all involved sectors with the same half hour cadence. The resulting light curves for HD 3191 were later cleaned from obvious outliers and shifted to
a zero mean before a combined analysis.
To support our assessment, we performed a
simultaneous least squares fit to the whole FFI dataset of four sine waves at the fundamental frequencies and their harmonics of the two newly found periods.
In particular, the sine wave frequencies that were kept fixed in the fit were 0.06215, 0.12430, 1.60527, and 3.21053 day$^{-1}$, 
which were derived from the above period analysis of only PDC Sectors 17 and 18. 
The modulation reconstructed using these sine waves mitigates the uncomplete sampling of the cycles. 
In Figs. \ref{qlporb} and \ref{qlprot},  we compare the 
first and second pair of sine components with the whole FFI dataset folded and averaged on the long and short period, respectively.
Here, it is remarkable that sine waves with the original frequencies are also able to provide a good average description of {\it TESS} photometry with sectors separated by
nearly half a year. This fact strongly suggests that the 16 and 0.6 day periods do indeed remain stable over timescales of at least several months.

\section{Discussion}

Here we adopt $M_V = -3.28$ and $BC=-2.67$ mag as the absolute magnitude and bolometric correction appropriate for
a B1 IV star, respectively \citep{2006MNRAS.371..185W, 2013A&A...550A..26N}. At the Gaia distance, this yields a star bolometric luminosity of
the $L_{\rm bol} \simeq 7.3 \times 10^{37}$ erg s$^{-1}$. Therefore we obtain the ratio $L_X/L_{\rm bol} \simeq  4 \times 10^{-7}$ that is consistent with the
'canonical' value typical of O-type and early B-type stars \citep{1997A&A...322..167B}. It is worth pointing out here that
binaries in B-type stars do not appear significantly X-ray brighter than single objects, thus leaving room for undiscovered companions \citep{2009A&A...506.1055N}.

The unexpected detection of an emission line of He II (4686 $\AA$) in Fig. \ref{LISA} immediately 
prompted us to consider a parallelism of HD 3191 with the binary system MWC 656,  which also exhibits such spectral feature arising in the accretion disk around its black hole companion
\citep{2014Natur.505..378C}.
We also have to recall here that both systems appear
 towards the direction of flaring gamma-ray sources, with which they could have a connection. Moreover, their respective spectral types are very similar, around B1-B2
 with a giant and subgiant luminosity class, except for the fact that MWC 656 also belongs to the Be star group with strong H$\alpha$ emission from a circumstellar envelope.
 
In any case, the strongest point of this work supporting the binary nature of HD 3191 is the clear detection of a nearly two-week period whose value fits well
within the typical range of orbital periods in high-mass X-ray binaries \citep{2006A&A...455.1165L}. At this point, we propose that $16.09 \pm 0.01$ d is likely
to correspond to the orbital cycle of HD 3191, which could harbor a compact companion eventually responsible for the He II emission as in MWC 656.

\subsection{Understanding the HD 3191 optical light curve}

\begin{figure}
\includegraphics[width=7.5cm, angle=-90]{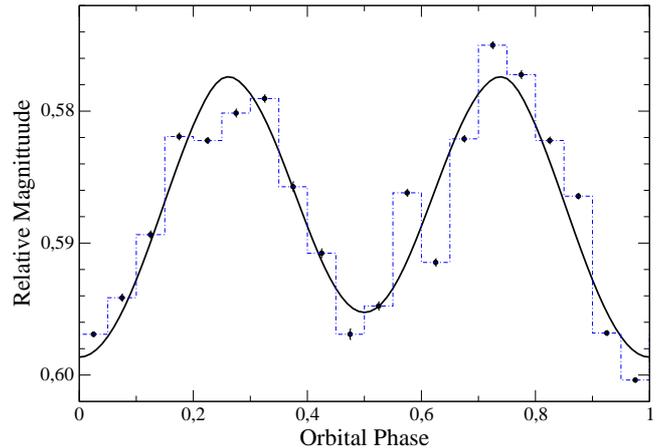}
%\imageii
\caption{   \label{fit}  Tentative attempt to model the average orbital light curve 
in terms of ellipsoidal variability due to the distorted primary star, as explained in the text.
Only PDC photometry of {\it TESS} Sectors 17 and 18 was 
%\LEt{ I'm not quite sure what you mean by "sed" here, possibly "set"?.} 
was used in the fit.
Bins of the 0.05 phase are shown.
}%% no full stop at the end
\end{figure}

To reinforce the binarity statement, we must pay further attention to the light curve appearance where one  expects
to see the typical double-wave pattern of ellipsoidal variability in massive X-ray binary systems. 
Hints of ellipsoidal variability were already suspected from OUJA photometry in Fig. \ref{folded_ouja}.
However, the TESS light curve in Fig.  \ref{tess_lc} % \ref{faseorb} 
shows that variability over the proposed orbital period 
is strongly distorted due to a second, short-term, sub-orbital period whose origin
has yet to be established. To mitigate this problem, we decided to proceed by averaging the TESS light curve with bins of 1/20 of the long-term period. The result is shown
in Fig. \ref{fit}, where we have also fitted a synthetic light curve based on a simplified version of the \citet{1971ApJ...166..605W}  code. 
The Fortran programme used was written by the corresponding author following the detailed equation formalism in \citet{2001icbs.book.....H}.
In Fig. \ref{fit}, the phase origin chosen in the previous section ensures that the zero orbital phase coincides with the inferior conjunction of the primary star.

Our custom code assumes, for simplicity, a circular orbit.
The physical effects taken into account include both gravity and limb darkening. Moreover, 
no reflection effects are considered in view of the low X-ray luminosity of the system.
The optical companion was kept  with fixed values of its mass  $M_1 = 15$ $M_{\odot}$ and effective temperature $T_1=25000$ K, 
appropriate for its spectral type and luminosity class.
In addition, the mass of the companion was set at a plausible value of $M_2 = 3$ $M_{\odot}$ and assumed to 
be a compact object (normalised surface potential $\Phi_{n, 2} \rightarrow \infty$) with an optical contribution from any accretion disk being neglected.
A summary of all relevant parameters used (both fixed and fitted) is condensed in Table \ref{param}. 

Despite the crudeness of our tentative fit, it manages to reproduce the general trend of the averaged light curve.
This fact encourages us to accept the  idea that the 16 d periodic
modulation in HD 3191 is of ellipsoidal origin.
For the parameters in Table \ref{param}, Roche lobe overflow would occur for normalised surface potential $\Phi_{n, 1} = 3.75$. 
The fitted value obtained turns out to be above, but not far from, this threshold (actually 2.7 standard deviations). 
Therefore, HD 3191 appears to be prone to possible mass-transfer episodes through the inner Lagrangian point that could eventually lead
to possible transient {\it Fermi} emission.

% 
%-------------------------------------------------------------
%                                             Simple A&A Table
%-------------------------------------------------------------
%
\begin{table}
\caption{Synthetic light curve parameters of HD 3191.}             % title of Table
\label{param}      % is used to refer this table in the text
\centering                          % used for centering table
\begin{tabular}{c c  }        % centered columns (4 columns)
\hline\hline                 % inserts double horizontal lines
Parameter & Value($^*$)   \\    % table heading 
\hline                        % inserts single horizontal line

Mass of the primary  & $M_1 = 15 $  $M_{\odot}$ \\            
Normalised surface potential & $\Phi_{n,1} =    4.65 \pm 0.22 $ \\ %& $-G(M_1+M_2)/2$ \\   
of the primary   & \\         
Effective temperature      & $T_1 = 25000$  K \\          
of the primary &   \\
Mass of the secondary & $M_2 = 3.0 $     $M_{\odot}$ \\             
%Surface potential($^{**}$)  & $\Omega_2 = 10$ & $-G(M_1+M_2)/2$  
%\\of the secondary &  & \\
Orbital inclination &  $35.8^{\circ} \pm 0.4^{\circ}$  \\
Orbital radius$^{**}$       & a=0.327 AU \\
\hline                                   %inserts single line
\end{tabular}
\tablefoottext{*}{Probable error is given for parameters being fitted. Others are fixed.}
\tablefoottext{**}{Orbital radius derived from $M_1$ and $M_2$,  
the orbital period and third Kepler's law.}
\end{table}

\begin{figure}
\includegraphics[width=7.5cm, angle=-90]{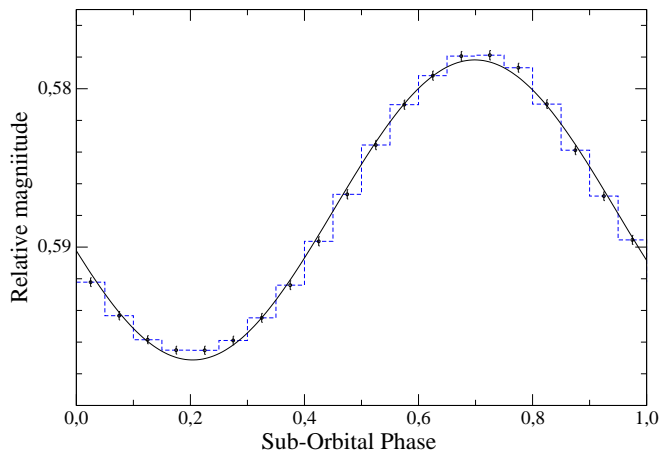}
%\imageii
\caption{   \label{faserot} Sinusoidal fit to the averaged light curve of HD 3191 folded on the short-term variability period. 
Only PDC photometry of {\it TESS} Sectors 17 and 18 is included.
Bins of the 0.05 phase are used.
}%% no full stop at the end
\end{figure}

Concerning the short-term, sub-orbital period in HD 3191, the most natural interpretation is in terms of rotation or pulsation. 
Its average behaviour is well represented by a pure sinusoidal function with an amplitude of 0.01 magnitudes, as shown in Fig. \ref{faserot}.
Among the two possible scenarios, 
we first consider the one
based on rotation because of the 0.6 d period persistency over several month scales,
its proposed detection in a few gamma-ray binaries \citep{2021arXiv210201971Z}, and
the abundant rotation evidences found by the
{\it TESS} observatory, not only in A-type stars, but also in B-type ones such as ours \citep{2019MNRAS.485.3457B}.
The fraction of rotational variables reported appears to be similar in both spectral types (about 40\%).
The rotational frequency in HD 3191 ($\nu_{\rm rot} = 1.6$ d$^{-1}$)  agrees well with the range of values within the B-type rotating variables.
A B1 IV is expected to have a surface gravity with $\log{g}=4.05$ \citep{2013A&A...550A..26N}, which implies a radius of about 6 $R_{\odot}$.
The observed sub-orbital period would then correspond to a rotation velocity close to 70\% of the break-up limit.
Actually, fast stellar rotation would not come as a surprise given the spectral classification of HD 3191 in the literature with  a suffix of nn used for rotationally broadened spectral lines.
In this scenario, the physical cause of the photometric modulation would be due to the presence of starspots, which would also
exist in early-type stars and not only in those of late-type, such as RS Canis Venaticorum systems.  These structures would arise as a result of
magnetic fields in the sub-surface convective zone of hot stars \citep{2009A&A...499..279C}, 
but their lifetime in early-type stars is not expected to last as long as our persistent sub-orbital period  \citep{2021MNRAS.500.2096R}.
 However, if large-scale magnetic fields are generated by extreme dynamo effects, then these magnetic structures could be stable on a much longer timescale,
 of approximately a year, at least for late B-type stars \citep{Cantiello_2019}.
In any case, the surface brightness distribution that would lead
to a nearly-perfect sinusoidal modulation in a rotation scenario is not a trivial issue.
For instance, the lack of synchronism between the rotation and the orbital period
would compromise the assumption of a Roche potential star shape implicit in codes of Wilson \& Devinney type.

The alternative interpretation of the sub-orbital period could be in terms of non-radial pulsation modes, known to be a long-lived stellar phenomenon.
Indeed, the luminosity and $\log{g}$ values quoted before HD 3191 are within the instability strip region as predicted by  \citet{2017A&A...597A..23G}.
The wide profile of absorption lines in its optical spectrum would then be reflecting
non-rotational line-broadening effects or macroturbulent broadening, also occurring in O stars and B supergiants according to these authors.
Here we speculate if enhanced pulsation, combined with even moderate eccentricity effects, could  provide the necessary kick-off to trigger
occasional mass-transfer episodes by Roche lobe overflow. 
%\LEt{ Single-sentence paragraphs are not allowed.}
Discriminating among rotation or pulsation scenarios could be addressed with accurate spectral modelling in the future.
  
\subsection{Is HD 3191 a runaway star?}

The modern {\it Gaia} EDR3 data can shed light on this issue already quoted in the introduction
which was first considered  by \citet{2011MNRAS.410..190T}. Using the latest astrometric information available
and the equation formalism given by \cite{1998A&A...331..949M}, it is possible to estimate the peculiar tangential 
velocity of the star $v_{\rm pec}$. This velocity component is the residual left when the proper motion due to Galactic rotation and
Sun motion with respect to the local standard of rest is subtracted from the observed heliocentric proper motion.
The stellar distance is assumed to be known in the calculation.
Runaway candidacy is considered to be likely if this residual exceeds about 42 km s$^{-1}$ plus the standard deviation of $v_{\rm pec}$.

In our case, we obtain $v_{\rm pec} = 5.9 \pm 0.5$ km s$^{-1}$ using {\it Gaia} EDR3. This result strongly argues against the runaway nature of HD3191 and agrees
with the fact that \citet{2011MNRAS.410..190T} could not confirm it as a runaway with historical {\it Hipparcos} data.
Nevertheless,
an estimate for the systemic radial velocity of the binary system is required to definitively determine whether or not it is a runaway. 
This motion along the line of sight could still provide a significant contribution to the modulus of the peculiar velocity with respect to the HD 3191 regional standard of rest.

\subsection{Search for the orbital period in {\it Fermi} data}

We have tried to find evidence for the proposed orbital period in the {\it Fermi} light curve of 4FGL J0035.8+6131, which could definitely set a positive identification.
According to our ephemeris, the two strong gamma-ray flares around MJD 57067 and MJD 57402 \citep{2018ApJ...862...83P} covered the orbital phase ranges from
0.39-0.42 and 0.18-0.26, respectively. Their overlap is marginal but suggestive at the $2\sigma$ level when considering the propagated uncertainty of 0.06 orbital cycles.

To gain further insight,
%For this purpose 
we obtained the {\it Fermi} LAT fluxes in the energy range  0.3-10 GeV where the gamma-ray source is better detected. They were  generated
%generated by aperture photometry 
with the latest version of the {\it Fermi} Science Tools (2.0.0 released on September 21, 2020). In order to be rigorous, instead of aperture photometry, we
used a likelihood analysis approach with  a $10^{\circ}$ radius around the position of 4FGL J0035+6131. The adopted  start and end times were
54682.65603 and 59167.51686 in the modified Julian day (MJD) scale with a time binning interval of 3 days. We also applied the perl script like\_lc.pl\footnote{
{\tt https://fermi.gsfc.nasa.gov/ssc/data/analysis/user/}
%and described in detail at
%{\tt https://fermi.gsfc.nasa.gov/ssc/data/analysis/user/like\_lc.txt}
}
 created by R. Corbet, using the current P8 R3 data, including the  gll\_iem\_v07 Galactic diffuse model and the iso\_P8R3\_SOURCE\_V3\_v1 isotropic spectral template, as well as nearby sources in the last gll\_psc\_v27.fit {\it Fermi} LAT catalogue. Finally, a simple power-law energy spectrum was fitted setting all parameters of the model fit to fixed, except the index and the integral of the target source. 
 
 \begin{figure}
\includegraphics[width=7.0cm, angle=-90]{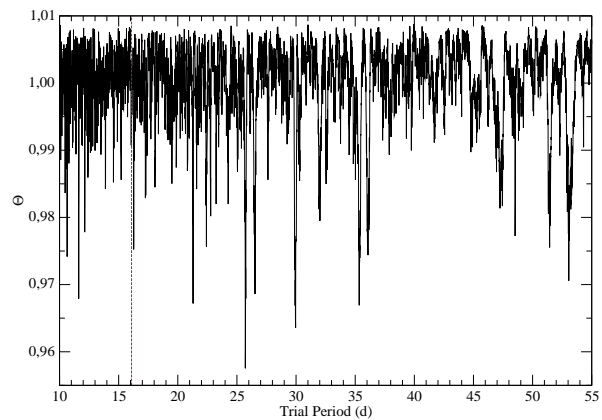}
%\imageii
\caption{   \label{fermiper} PDM analysis of the {\it Fermi}-LAT curve of 4FGL J0035.8+6131 in the 0.3-10 GeV energy range.
The red vertical line marks the expected location of the orbital period of HD 3191 where no clear signal is present.
}%% no full stop at the end
\end{figure}

The resulting light curve turned out to be rather noisy, except at times of previously known flares. The result of a PDM periodogram  
is displayed in Fig. \ref{fermiper}. Unfortunately, the signature of the sixteen day orbital period is not confidently detected with the present data set.
Several of the deepest minima in the periodogram are of instrumental nature and related to the 53 d precession period of the satellite and its
harmonics. Yet, this does not strictly rule out the possible association with HD 3191 as the suspected gamma-ray emission is clearly not a steady one.

\section{Conclusions}

An extended observational study of the gamma-ray source 4FGL J0035.8+6131 focused on its candidate counterpart HD 3191 has been presented.
Our main conclusions can be summarized as follows:

\begin{enumerate}

\item
We have reported strong photometric evidence on the binary nature of the early-type star HD 3191. This finding supports its previously proposed classification
as an X-ray binary system. Two clear periodic modulations, close to 16 d and 0.6 d,  are evident in the HD 3191 optical light curve. Their most likely interpretation is in terms of the
orbital and rotational or pulsation cycle of the optical companion, respectively.

\item Spectroscopic observations have revealed an interesting parallelism between HD 3191 and the black hole X-ray binary MWC 656. Both are showing a HeII
4686 \AA\ emission line and have been tentatively associated
with different flaring gamma-ray sources. In this context, the possible connection of HD 3191 with the {\it Fermi} source 4FGL J0035.8+6131, instead of a blazar
identification, remains a very conceivable one. The helium emission in HD 3191 could be related to a compact companion as in MWC 656. An extensive radial velocity
monitoring is necessary to confirm or rule out this suspicion.

\item Photometric modelling tentatively suggests that the system is below, but close to Roche lobe overflow.  Enhanced mass-transfer episodes, triggered
by  pulsation effects or a slightly eccentric orbit, could eventually play a role in occasional high-energy emission events.

\item A few km s$^{-1}$ value has been obtained for the HD 3191 peculiar tangential velocity with respect to its environments, thus suggesting that we are not dealing with
a runaway system.

\item A search for the orbital period in the 4FGL J0035.8+6131 light curve has produced negative results that prevent, so far, an unambiguous association
with the HD 3191 system. 

\end{enumerate}

The confirmation of HD 3191 as  a binary is possibly the most important contribution of this paper.
This is because it now opens the opportunity to attempt the orbital period detection in future improved {\it Fermi}-LAT data sets.

\begin{acknowledgements}
We thank an anonymous referee for helping to improve this paper.
This work was mainly supported  by grant PID2019-105510GB-C32 / AEI / 10.13039/501100011033 from 
State Agency for Research of the Spanish Ministry of Science and Innovation 
 entitled {\it High energy sources with outflows at different scales: observation of galactic sources}. We also acknowledge support
 by Consejer\'{\i}a de Econom\'{\i}a, Innovaci\'on, Ciencia y Empleo of Junta de Andaluc\'{\i}a as research group FQM- 322, as well as FEDER funds.
JMP acknowledges financial support from the State Agency for Research of the Spanish Ministry of Science and Innovation
under grant PID2019-105510GB-C31 and through the Unit of Excellence Mar\'{\i}a de Maeztu 2020-2023 award to the
Institute of Cosmos Sciences (CEX2019-000918-M), and by the Catalan DEC grant 2017 SGR 643.
 This paper includes data collected with the TESS mission, obtained from the MAST data archive at the Space Telescope Science Institute (STScI). Funding for the TESS mission is provided by the NASA Explorer Programme. STScI is operated by the Association of Universities for Research in Astronomy, Inc., under NASA contract NAS 5-26555.
 This research has made use of the SIMBAD database,
operated at CDS, Strasbourg, France.
This work has also made use of data from the European Space Agency (ESA) mission
{\it Gaia} (\url{https://www.cosmos.esa.int/gaia}), processed by the {\it Gaia}
Data Processing and Analysis Consortium (DPAC,
\url{https://www.cosmos.esa.int/web/gaia/dpac/consortium}). Funding for the DPAC
has been provided by national institutions, in particular the institutions
participating in the {\it Gaia} Multilateral Agreement.
\end{acknowledgements}

\bibliographystyle{aa} % style aa.bst
\bibliography{references}

\end{document}